\documentclass[journal]{IEEEtran}
\pdfoutput=1
\usepackage{graphicx}
\usepackage{amsfonts}
\usepackage{amssymb}
\usepackage[cmex10]{amsmath}
\begin{document}
\title{A Class of Random Binary Waveforms\\ With Impulse-Like Autocorrelation} 
\author{W.~J.~Szajnowski~\IEEEmembership{}
        {}% <-this % stops a space
\thanks{W. J. Szajnowski is a Visiting Professor at %VSSP 
	Centre for Vision, Speech and Signal Processing, University of Surrey, Guildford GU2 7XD, U.K., e-mail: w.j.szajnowski@surrey.ac.uk}% <-this % stops a space
\thanks{}
}

\markboth{ }%
{Shell \MakeLowercase{\textit{et al.}}: Bare Demo of IEEEtran.cls for Journals}

\maketitle

\begin{abstract}
\boldmath
 It is shown that a random binary process with impulse-like autocorrelation  can be generated by randomizing the length of symbols occurring in a random Bernoulli process. Such randomization is achieved by random (or judiciously designed irregular) sampling of the output of a source that supplies a symmetric random binary waveform. One practical configuration implementing the proposed technique is a linear-feedback shift register driven by a spread-period clock generator.
 
 \end{abstract}


\IEEEpeerreviewmaketitle

\section{Introduction}
\IEEEPARstart{T}{he} generation of random binary waveforms with specified correlation properties is of considerable theoretical and practical interest in many fields, such as system identification, remote sensing, communications, navigation and radar. In most practical applications, there is a requirement to employ a random binary waveform whose autocorrelation function has a shape that can approximate, in some manner, an impulse function [1]--[4].

A basic model of a random binary waveform is a Bernoulli sequence that comprises realizations of independent and identically distributed random binary variables. However, in many cases, it is more convenient to exploit a {\emph {quasi-random}} or {\emph {pseudorandom}} binary sequence (PRBS) to obtain a much faster convergence of empirical (time) averages to corresponding (ensemble) means. Such an approach is similar to replacing a standard Monte Carlo (MC) method by a much more efficient quasi-Monte Carlo (QMC) technique that employs non-random numbers with specific characteristics [5].

In practical applications, a pseudorandom sequence of binary symbols, e.g., generated by some algebraic method, has to be converted into a physical {\emph {base-band}} waveform in which each of the two symbols is represented by a distinct voltage level. Additionally, a binary waveform with prescribed correlation properties can be used to modulate a radio-frequency carrier. For example, in low probability of
intercept/low probability of detection (LPI/LPD) radar, the phase of a coherent carrier is modulated by a PRBS to spread the spectrum of the transmitted signal [6], [7].

In some applications, such as collision avoidance/obstacle detection, altimetry, autonomous navigation etc., many similar radar systems should be capable of operating in the same region and sharing the same wide frequency band. In order to avoid mutual interference, each system should use a distinct waveform, preferably orthogonal to the waveforms employed by all other systems. This can be achieved by randomization, i.e. a single radar waveform can be exploited to produce a number of mutually uncorrelated replicas by modifying in a (pseudo)random fashion a suitably chosen parameter of the underlying waveform.

The main objective of this paper is to analyze some specific statistical properties of a random binary process and to develop a practical technique to generate a random binary waveform that exhibits an impulse-like autocorrelation function.

\section{A Random Binary Process}
A random binary process $V(t)$ with unit amplitude may be represented as [8]
\begin{equation}
V(t) =\, (-1)^i \mspace{1mu} \Theta; 
\quad t_i <t <t_{i+1},\quad i=0,\pm 1, \ldots 
\end{equation}
where the zero crossing points $\{t_i\}$ are assumed to constitute a stationary and ergodic point process. The random parameter $\Theta$ may take on one of two values, $+1$ or $-1$, with equal probability. The average zero-crossing rate is assumed to be finite and equal to ${n}_0$.

The autocorrelation function $R_V(\tau)$ of a binary process $V(t)$ is defined by
\begin{equation}
 R_V(\tau) \triangleq \, \mathrm{E}\{V(t)V(t+\tau)\}
 \end{equation}
where $\mathrm{E}\{\cdot\}$ denotes statistical expectation. 
By definition, $\mathrm{E}\{V(t)\}=0$, and 
$R_V(0)=\mathrm{E}\{V^2(t)\}=1$.

The autocorrelation function $R_V(\tau)$ is right and left differentiable, and has a cusp at the origin. Furthermore,
\begin{equation}
R_V(\tau) \,\ge \,
 \left(1 - 2\mspace{1mu}{n}_0 |\tau| \right)\mspace{-2mu},
\quad |\tau| < \infty
\end{equation}
i.e. the autocorrelation function remains above its tangent drawn at the origin. Hence, the width of the autocorrelation function $R_V(\tau)$ depends on the mean zero-crossing rate $n_0$, and the function $R_V(\tau)$ becomes narrower as the rate $n_0$ is increasing.

The mean zero-crossing rate, $n_0$, of a random binary process $V(t)$ can be determined from
\begin{equation}
n_0 =
- \frac {1}{2} \lim_{\tau \to 0^+} R'_V(\tau) =
\frac {1}{2} \lim_{\tau \to 0^-} R'_V(\tau).
\end{equation}
Since
\begin{equation}
R'_V(0^+)  = -2\mspace{1mu}{n}_0
\quad {\mbox {and}} \quad
R'_V(0^-)  = 2\mspace{1mu}{n}_0
\end{equation}
the second derivative $R''_V(0)$ at the origin of the autocorrelation function $R_V(\tau)$ can be represented by $-4\mspace{1mu}{n}_0 \mspace{1mu} \delta(\tau)$, where $\delta(\cdot)$ is the Dirac delta function. 

The value, $-4\mspace{1mu}{n}_0 \mspace{1mu} \delta(\tau)$, of $R''_V(0)$ can also be obtained from the relationship, 
$R''_V(0) = -\mathrm{E}\{[V'(t)]^2\}$, where $V'(t)$ is a symbolic random process comprising positive and negative impulse functions, $2\delta(\cdot)$ and $-2\delta(\cdot)$, located at corresponding zero upcrossings and downcrossings of the underlying binary process $V(\tau)$.

The {\emph {angular}} mean-square (ms) bandwidth, $\beta^2_V$, of a random binary process $V(t)$ can be determined from 
\begin{eqnarray}
\beta^2_V \mspace{-7mu}
&\triangleq& \mspace{-5mu}
\frac {\int_{-\infty}^{\infty}\mspace{-2mu} \omega^2 S_V(\omega) \, d\mspace{1mu}\omega} {\int_{-\infty}^{\infty} S_V(\omega)\, 
	d\mspace{1mu}\omega} \nonumber \\
&=& \mspace{-5mu}
 \frac {-R''_V(0)} {R_V(0)} = -{R''_V(0)} 
\,=\, 4\mspace{1mu}{n}_0 \mspace{1mu} \delta(\tau)
\end{eqnarray}
where $\omega$ is the angular frequency, and $S_V(\omega)$ is the power spectrum density (psd) of $V(t)$. Hence, the ms bandwidth $\beta^2_V$ of any random binary process $V(t)$ is infinite.

It should be pointed out that the \emph{theoretically} infinite ms bandwidth $\beta^2_V$ of a random binary process $V(t)$ is not affected by the value of its zero-crossing rate $n_0$. However, a greater rate $n_0$ will result in a smaller width of the autocorrelation function $R_{V}(\tau)$. Since the error of time-delay estimation depends directly on the ms bandwidth $\beta^2_V$, and not on the rate $n_0$, conventional correlation processing cannot lead to an optimal solution of the problem of \emph{joint} detection/localization of binary waveforms.   

In practical applications, a random binary process is generated by a physical (power-limited) source so that the switching times between the two levels of the process are greater than zero. Consequently, the cusp (at $\tau=0$) of the resulting autocorrelation function will locally be replaced by a differentiable shape of parabolic type, and the resulting ms bandwidth will always be finite.

\subsection{Random Telegraph Signal}
One example of a random binary process (1) is the \emph{random telegraph signal}, defined as a binary waveform with transitions between the levels $\pm 1$, occurring at random time instants $\{t_i\}$.
In this case, the intervals $\{|t_{i+1} - t_i|\}$ are all independent random variables having the same exponential distribution [2]. 

If the mean zero-crossing rate of the random telegraph signal is $n_0$, then its autocorrelation function is of the form
\begin{equation}
R_V(\tau)  \,=\, \exp(-2\mspace{1mu} n_0 |\tau|). 
\end{equation} 
As seen,
\begin{equation}
R'_V(0^-)  \,=\, -R'_V(0^+)  \,=\, 2\mspace{1mu}{n}_0
\end{equation}
so that the zero-crossing rate of the random telegraph signal is in agreement with the general formula (4).
 
The autocorrelation function of the random telegraph signal is shown in Fig.\,1\,a.

The power spectrum density $S_V(\omega)$ of the random telegraph signal has a Lorentzian shape, given by
\begin{equation}
 S_V(\omega) \,=\,
  \mathcal{F}\{R_V(\tau)\} \,=\,
   \frac {4\mspace{1mu} n_0} {\omega^2+4\mspace{1mu} n_0^2} \mspace{2mu}. 
\end{equation} 
where $\mathcal{F}\{\cdot \}$ denotes the Fourier transform.

\begin{figure}[] %[b]
	\centering
	\includegraphics[width=9cm]{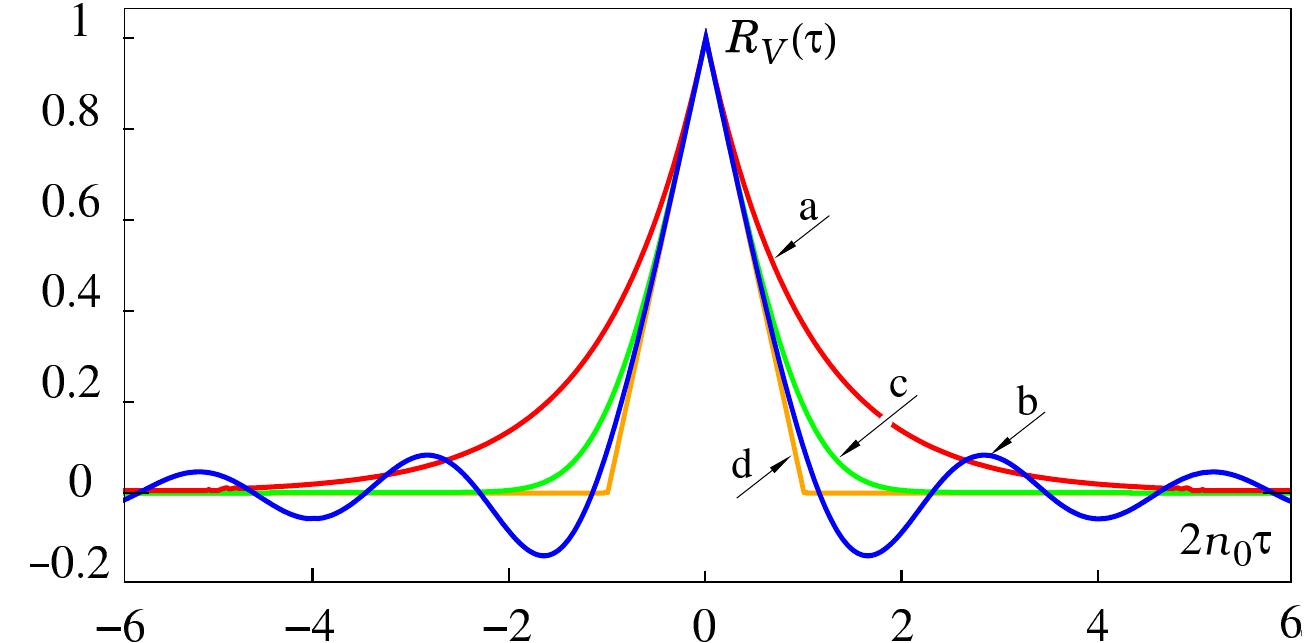}
	\vspace{-0.6cm}
	\caption{Autocorrelation function $R_V(\tau)$ of several random binary processes: (a) random telegraph signal, hard-limited Gaussian noise with (b) uniform and (c) Gaussian power spectrum density, (d) constant-chip Bernoulli process.}
\end{figure}

\subsection{Hard-Limited Gaussian Noise}
An important class of random binary processes is obtained by applying 'hard limiting' (a signum operation) to a real-valued zero-mean Gaussian process $X(t)$, i.e.
\begin{equation}
V(t)\,=\,
\mathrm{sgn}[X(t)]\, = \,
\left\{
\begin{array}{r@{\,\,\quad}l}\!\!+1, & {X(t)>0}\\
\!\!0, & {X(t)=0}\\
\!\!-1, & {X(t)<0} \, .
\end{array} \right . 
\end{equation}
It is assumed that $X(t)$ is a stationary and ergodic random process with normalised autocorrelation function  $r_X(\tau)\triangleq R_X(\tau)/R_X(0)$. 
The Gaussian process $X(t)$ is also assumed to have continuously differentiable realizations (sample paths) for which zero crossings are easily defined. 

The normalized autocorrelation function, $r_X(\tau)$, is assumed to be twice differentiable, so that the \emph {angular} ms bandwidth $\beta^2_X$ is equal to $-r''_X(0)$. From Rice's formula, it follows that the mean zero-crossing rate of $X(t)$ is equal to $\beta_X/\pi$ [9]. By construction, zero crossings of the process $X(t)$ and those of $V(t)$ will coincide; hence, $n_0 = \beta_X/\pi$. 

As a consequence of (3), the width of the autocorrelation function $R_V(\tau)$ becomes narrower as the rms bandwidth $\beta_X$ of the underlying Gaussian process $X(t)$ is increasing. However, the rms bandwidth $\beta_V$ of the resulting binary process $V(t)$ will remain infinite, irrespective of the value of $\beta_X$.

The autocorrelation function $R_V(\tau)$ of a random binary process $V(t)$, obtained from the underlying Gaussian process $X(t)$, is given by 
\begin{equation}
R_V(\tau)  = \frac {2}{\pi} \arcsin[r_X(\tau)].
\end{equation}
The above relationship is often referred to as the arcsine law or Van Vleck's formula [10], [11].

Fig.\,1 shows plots of the autocorrelation function $R_V(\tau)$ of a random binary process $V(t)$ generated by Gaussian noise with:\\
1. psd uniform in the interval $|\omega|<W$ and normalized autocorrelation function
\begin{equation*}
r_X(\tau) \,=\,  {\sin( W \tau )}/{(W\tau)}, \quad W = \sqrt{3} \mspace{1mu} \pi n_0, \quad \mbox{(Fig.\,1\,b)}.
\end{equation*}
2. Gaussian psd and normalized autocorrelation function
\begin{equation} 	
r_X(\tau) \,=\, \exp(-\pi^2 n^2_0\mspace{1mu} \tau^2\mspace{-1mu} /2),
\quad \mbox{(Fig.\,1\,c)}.
\end{equation}

\subsection{A Synchronous Random Binary Process}
In a \emph{synchronous} random binary process, the distance between consecutive zero-crossing points is an integer multiple of a constant time interval. The process is specified jointly by a sequence of independent and identically distributed binary random variables and a constant interval of each symbol duration; this interval is often referred to as the clock or  \emph{chip} period. Consequently, in the following, such a process will be referred to as a {\emph {constant-chip Bernoulli process}}.

The autocorrelation function, $R_{B}(\tau)$, of a constant-chip Bernoulli process $B(t)$ is of the form [2]
\begin{equation}
R_{B}(\tau) \,=\, 
{\Lambda}\mspace{1mu} (\tau; t_C) 
\end{equation}
where $t_C$ is the chip period, and 
\begin{equation}
{\Lambda}\mspace{1mu} (\tau; \zeta) 
\,\, \triangleq \,\,
\left\{
\begin{array}{l@{\,\,\,\,\,\,\,\,\,}l} 
{\mspace{-4mu}   1- {|\tau|}/{\zeta}  ,} 
&{|\tau|\leq \zeta}\\
{\mspace{-4mu}0,} & {|\tau| > \zeta}\, .
\end{array} \right .
\end{equation}
In accordance with (4), the mean zero-crossing rate $n_0$ of the process $B(t)$ is given by
$$
n_0 = 1/(2\mspace{2mu}t_C)
$$ 
i.e. $n_0$ is equal to a half of the chip rate.

The autocorrelation function of a constant-chip Bernoulli process is shown in Fig.\,1\,d.

The power spectrum density of a constant-chip Bernoulli process can be determined from
\begin{equation}
S_{B}(\omega)  \, \triangleq \,
\mathcal{F}\{{\Lambda}\mspace{1mu} (\tau; t_C) \} \,=\, 
\frac {4\mspace{1mu}\sin^2(\omega\mspace{1mu} t_C/2)}{\omega^2 \mspace{1mu}t_C}\, .
\end{equation}
The \emph{theoretically} infinite ms bandwidth of a constant-chip Bernoulli process results from the instantaneous switching of its values between the two levels. While the ms bandwidth of the process is not affected by the chip period $t_C$, the \emph{width} of the autocorrelation function $R_B(\tau)$ is directly proportional to the value of $t_C$.

\subsection{Zero-Crossing Interferogram of a Random Binary Process}
Consider a single realization, a binary waveform $v(t)$, of an underlying random binary process $V(t)$. Let $\{t_i\}$ be the time instants of observed zero crossings and attach to each $t_i$ a  binary trajectory $v_i(\tau) \triangleq v(t_i+\tau)$, where $\tau$ denotes relative time. It should be noted that each binary trajectory $v_i(\tau)$ is simply a time-shifted copy of the entire waveform $v(t)$. Since, by construction, the time instants $\{t_i\}$ will all have collapsed onto a single point $\tau = 0$, the corresponding binary trajectories will share the same origin of relative time $\tau$.

Binary trajectories $\{v_i(\tau)\}$ can be used to determine an empirical zero-crossing interferogram [12], [13]
\begin{equation}
\widehat{\mathcal{D}}_V(\tau)\, \triangleq\,
\frac{1}{n_c} \mspace{-2mu}
\sum_{i = 1}^{n_c} (-1)^{\rho_i}v_i(\tau)
\end{equation}
where $n_c$ is the number of observed zero crossings; $\rho_i=0$ for an upcrossing, and $\rho_i=1$ for a downcrossing at $t_i$.

It can be shown that 
\begin{equation} 
\lim_{n_c \to \infty}\widehat{\mathcal{D}}_V(\tau) \,\,=\, 
%\gamma \mspace{1mu} \mathrm{E}\{B'(t)\mspace{1mu}B(t+\tau)\}
-\frac {1}{2\mspace{1mu}n_0}\mspace{1mu} R'_V(\tau)
\end{equation}
where $n_0$ is the zero-crossing rate [12].

For illustration purposes, Fig.\,2 shows five superimposed empirical zero-crossing interferrograms of a random binary process obtained by hard-limiting of Gaussian noise with Gaussian  autocorrelation function given by (12). Each empirical interferrogram $\widehat{\mathcal{D}}_V(\tau)$ is the result of averaging of $n_c\!=\!256$ binary trajectories (sample paths) associated with consecutive zero crossings. The $95$\% confidence interval has been determined for the mean value of $\widehat{\mathcal{D}}_V(\tau)$,
\begin{equation*}
\mathrm{E}\{\widehat{\mathcal{D}}_V(\tau)\} \, = \,
\lim_{n_c \to \infty}\widehat{\mathcal{D}}_V(\tau) \,=\, 
- \frac {\pi n_0 \tau \exp(-\pi^2 n_0^2 \tau^2/2)}
{\sqrt{1 - \exp(-\pi^2 n_0^2 \tau^2)}}.
\end{equation*}

 It should be pointed out that the empirical zero-crossing interferogram $\widehat{\mathcal{D}}_V(\tau)$, resulting from a {\emph {linear}} operation (16), i.e. averaging conditioned on zero crossings, represents correlation properties of a random binary process $V(t)$ under study.

\begin{figure}%[b]
	\begin{center}
		\includegraphics[width=7.8cm]{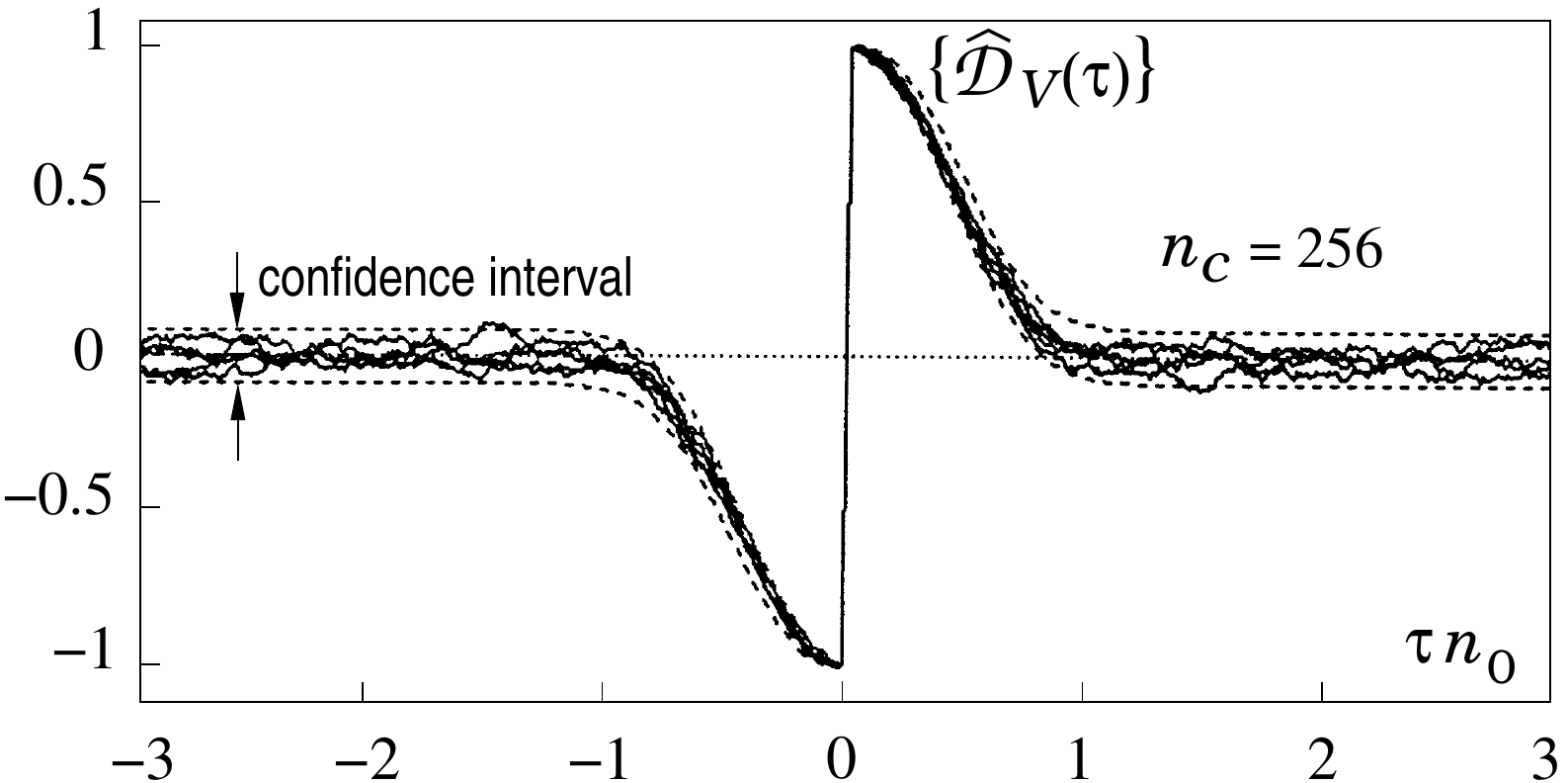}
	\end{center}
	\vspace{-0.4cm}
	\caption{Superimposed empirical zero-crossing interferograms of a random binary process obtained by hard-limiting of Gaussian noise.}
\end{figure}

\subsection{A Model of Optimal Autocorrelation Function}
A simple model of the optimal autocorrelation function $R_V(\tau)$ of a random binary process $V(t)$ can be obtained by replacing parts of the two linear slopes of the triangular function (14) by segments of a suitably chosen parabola $q(\tau;\alpha)$. 

Such a construction of the proposed model is shown schematically in Fig.\,3\,a.
For mathematical convenience, and without any loss of generality, it has been assumed that $t_C=1$.

When $\tau > 0$, the two conditions of a smooth transition at the point $\tau = 1 - \alpha$:
\begin{equation}
q(1\mspace{-1mu}-\mspace{-1mu}\alpha;\alpha) \,=\, \alpha \,\quad\mbox{and}\,\quad
\frac {\partial\mspace{1mu} q(\tau;\alpha) }
{\partial\mspace{1mu} \tau} \bigg|_{\tau=1-\alpha}\mspace{-4mu}=\, -1 
\end{equation}
are satisfied by the left branch of the parabola
\begin{equation}
q(\tau;\alpha) \,=\, 
{ (
	\mspace{1mu} \tau \mspace{-1mu}-\mspace{-1mu}1 \mspace{-1mu}-\mspace{-1mu}\alpha
	 )^2}\mspace{-2mu}/
{(4\mspace{1mu} \alpha)}; 
\quad 0\mspace{-1mu} \leq \mspace{-1mu}\alpha \mspace{-1mu}\leq\mspace{-1mu} 1, 
\quad \tau\mspace{-1mu} >\mspace{-1mu} 0 \mspace{1mu}.
\end{equation}

The above construction has resulted in the model function 
\begin{equation}
\widetilde{\Lambda}(\tau;\alpha) \mspace{-1mu}=\mspace{-1mu}
\left\{
\begin{array}{r@{\,\,\quad}l}
\mspace{-10mu}1\mspace{-1mu}-\mspace{-1mu}|\tau|, & {|\tau|\mspace{-1mu} <\mspace{-1mu} 1\mspace{-1mu}-\mspace{-1mu}\alpha}\\
\mspace{-10mu}%\frac 
{\left (
	\mspace{1mu} |\tau|\mspace{-1mu}-\mspace{-1mu}1 \mspace{-1mu}-\mspace{-1mu}\alpha
	\right )^2}\mspace{-2mu}/
{(4\mspace{1mu} \alpha)},
%(1\mspace{-1mu}-\mspace{-1mu}|\tau| \mspace{-1mu}+\mspace{-1mu}\alpha)^2, 
& {1\mspace{-1mu}-\mspace{-1mu}\alpha \mspace{-1mu}<\mspace{-1mu} |\tau| \mspace{-1mu}< \mspace{-1mu}1\mspace{-1mu}+\mspace{-1mu}\alpha}\\
\mspace{-10mu}0, & {|\tau|\mspace{-1mu} >\mspace{-1mu} 1\mspace{-1mu}+\mspace{-1mu}\alpha} \, .
\end{array} \right . 
\end{equation}
By construction, the function $\widetilde{\Lambda}(\tau;\alpha)$ is positive, has a cusp at $\tau=0$, and is decreasing monotonically from its maximum value $\widetilde{\Lambda}(0;\alpha)=1$ to zero.

Since the function $\widetilde{\Lambda}(\tau;\alpha)$ satisfies the conditions [14], [15]:\\
(a)\,\, $\widetilde{\Lambda}(0;\alpha) = 1$\\
(b)\,\, $\widetilde{\Lambda}(\tau;\alpha) = \widetilde{\Lambda}(-\tau;\alpha)$\\
(c)\,\, 
$\widetilde{\Lambda}^2(\tau;\alpha) \,\leq \,
\frac {1}{2}   [1 + \widetilde{\Lambda}(2\mspace{1mu}\tau;\alpha)]$\\
(d)\,\, $\lim_{\tau \to \infty}\widetilde{\Lambda}(\tau;\alpha) = 0$\\
the function $\widetilde{\Lambda}(\tau;\alpha)$ can be considered as a realizable model of an autocorrelation function of a hypothetical random binary process.

The model function (20) is positive, has no side-lobes, and can be made very narrow by increasing the zero-crossing rate. Therefore, in practical applications, the function $\widetilde{\Lambda}(\tau;\alpha)$ may provide a good approximation of an impulse function $\delta(\tau)$.

 \begin{figure}%[b]
	\begin{center}
		\includegraphics[width=7.5cm]{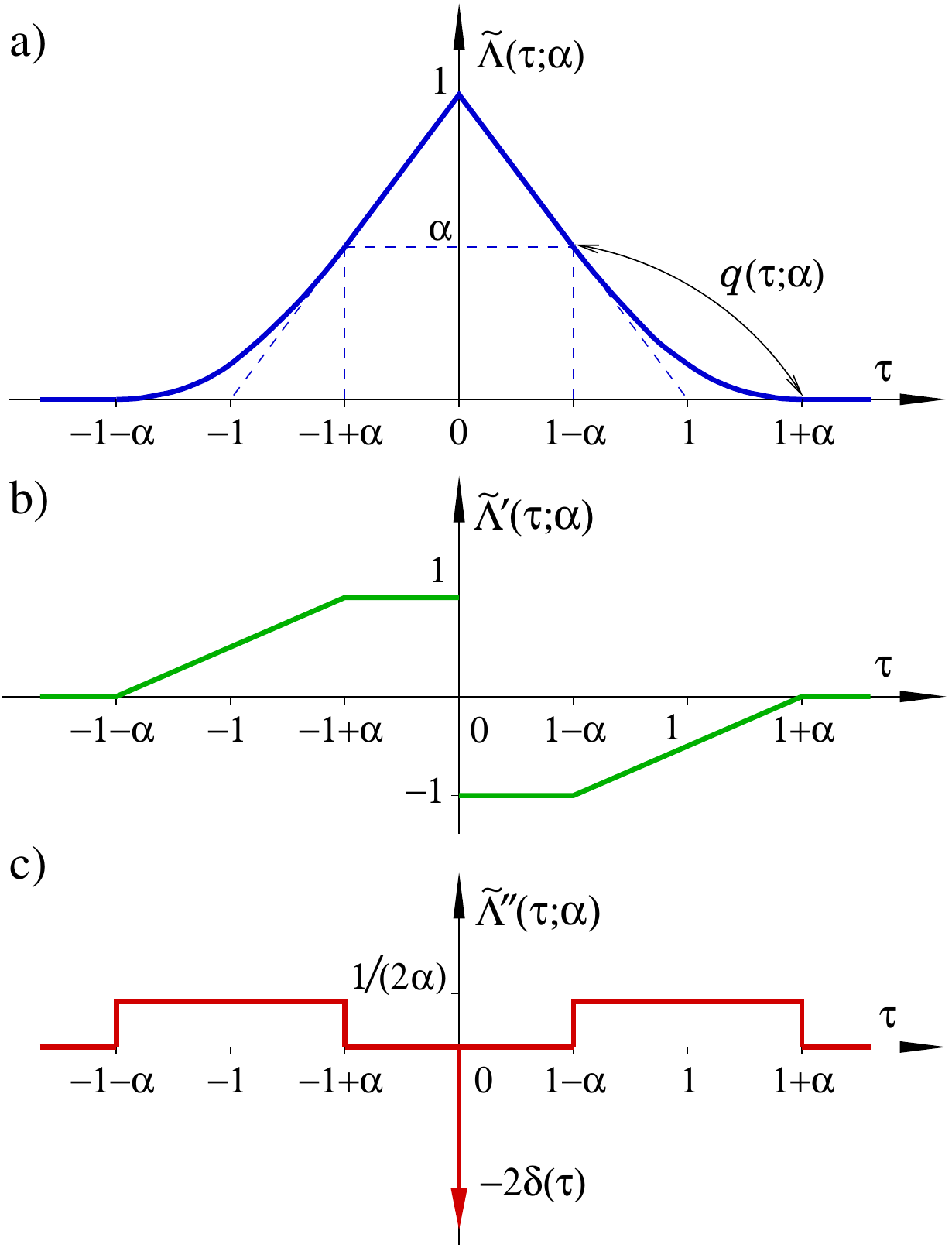}
	\end{center}
	\vspace{-0.3cm}
	\caption{Autocorrelation function $\widetilde{\Lambda}(\tau;\alpha)$ and its two derivatives of a hypothetical random binary process.}
\end{figure}
The above analysis has led to two basic questions:\\
1. Is there a technique capable of generating an asynchronous random binary process with the autocorrelation function (20);\\
2. Can such a technique, if found, be employed to generate asynchronous random binary processes with other shapes of optimal autocorrelation function.\\
The answers to the above questions are presented below.

\section{Bernoulli Processes With Random Duration of Binary Symbols}
Consider a single realization, a binary waveform $b(t)$, of an underlying constant-chip Bernoulli random binary process $B(t)$ with the autocorrelation function $R_B(\tau)$ shown in Fig.\,4\,a. Let $\{t_i\}$ be the time instants of observed zero crossings and attach to each $t_i$ a  binary trajectory $b_i(\tau) \triangleq b(t_i+\tau)$, where $\tau$ denotes relative time. An empirical zero-crossing interferogram assumes the form
\begin{equation*}
\widehat{\mathcal{D}}_B(\tau)\, \triangleq\,
\frac{1}{n_c} \mspace{-2mu}
\sum_{i = 1}^{n_c} (-1)^{\rho_i}b_i(\tau)
\end{equation*}
where $n_c$ is the number of observed zero crossings; $\rho_i=0$ for an upcrossing, and $\rho_i=1$ for a downcrossing at $t_i$.

In this case,
\begin{equation} 
\mathcal{D}_B(\tau)\,\triangleq \,
\mathrm{E}\{\widehat{\mathcal{D}}_B(\tau)\} \, = \,
\lim_{n_c \to \infty}\widehat{\mathcal{D}}_B(\tau) \,=\,
-t_C\mspace{1mu} R'_B(\tau)
\end{equation}
where $t_C$ is the chip period, and $R'_B(\tau)$ is the derivative of the autocorrelation function $R_B(\tau)$.

The derivative $R'_B(\tau)$ can be expressed as 
\begin{eqnarray}
R'_B(\tau) &\mspace{-8mu}=\mspace{-8mu}&
\left\{
\begin{array}{l@{\,\,\,\,\quad}l} 
{\mspace{-9mu}  [- u(\tau) + u(\tau-t_C)]/t_C ,} &{\tau > 0 }\\
{\mspace{-9mu}	[+ u(-\tau) - u(-\tau-t_C)]/t_C ,} &  {\tau < 0}
\end{array} \right .\nonumber \\
&\mspace{-8mu}=\mspace{-8mu}&{\mspace{3mu} \mathrm {sgn}(\tau)\mspace{1mu}
	\left[- u(|\tau|) + u(|\tau|-t_C) \right]/t_C} 
\end{eqnarray}
where $u(\tau)$ is the unit step function.

The derivative $R'_B(\tau)$ of the autocorrelation function $R_B(\tau)$ of a constant-chip Bernoulli process $B(t)$ is shown in Fig.\,4\,b.

\begin{figure}%[b]
	\begin{center}
		\includegraphics[width=6.5cm]{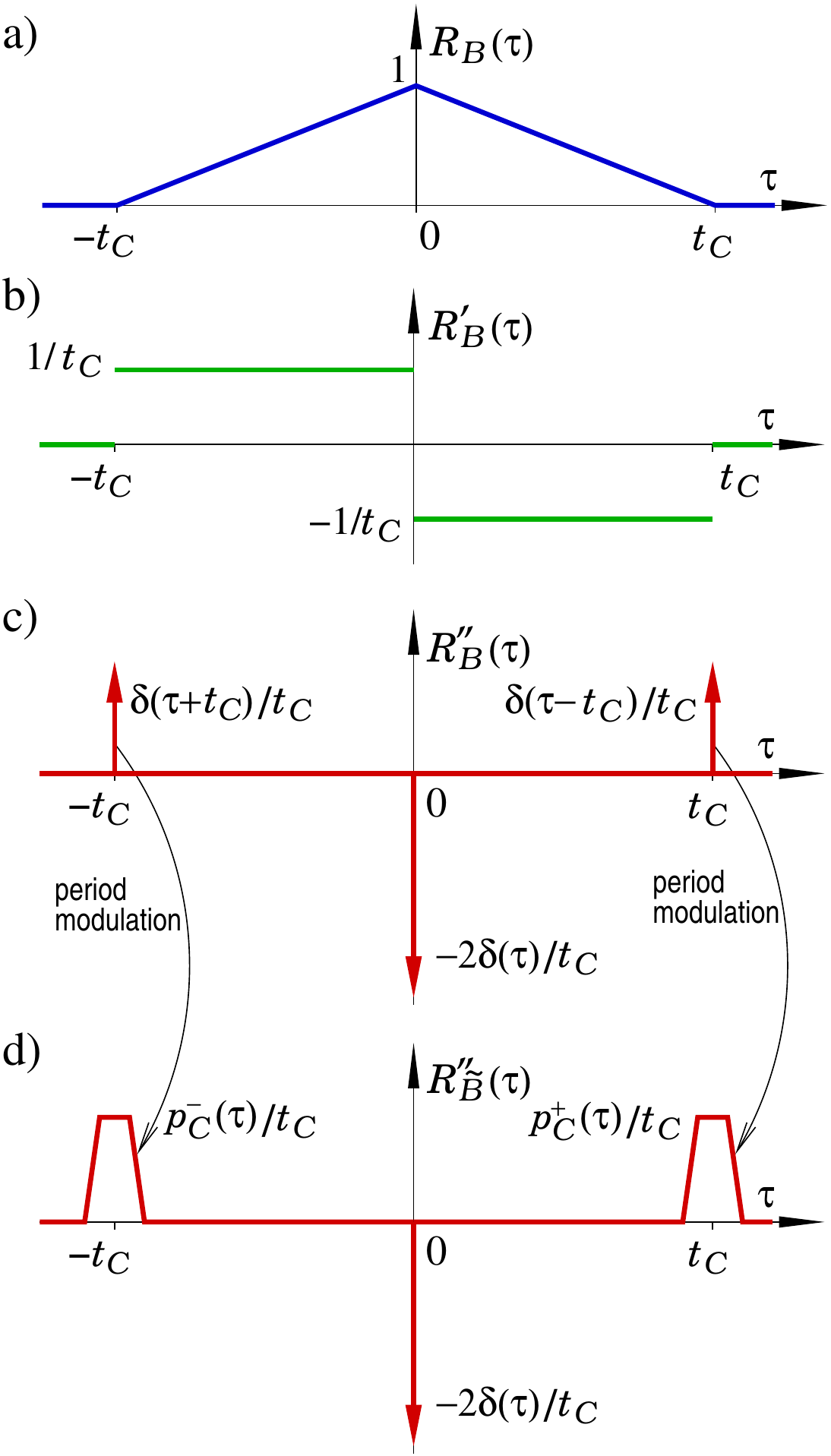}
	\end{center}
	\vspace{-0.3cm}
	\caption{Autocorrelation function $R_B(\tau)$ and its two derivatives of a constant-chip Bernoulli process.}
\end{figure}

\subsection{A Random-Chip Bernoulli Process}
Suppose now that the chip period $t_C$ is being varied in a random fashion with respect to its nominal value. More specifically, assume that the chip period is a random variable $C$ with the mean value $t_C$ and
probability density function 
\begin{equation} 
p^+_C(\tau)  =  p_C(\tau-t_C) 
\end{equation}
where $p_C(\tau)$ is a unimodal and even function of $\tau$ and has a finite support 
$(-\alpha \mspace{1mu} t_C , 
\alpha \mspace{1mu} t_C )$, where $0 \leq \alpha \leq 1$.

Such random modulation of the chip period will produce an asynchronous random binary process ${\widetilde{B}}(\tau)$, and the corresponding zero-crossing interferogram 
$\widehat{\mathcal{D}}_{\widetilde{B}}(\tau)$ will converge to the new mean,
\begin{equation} 
\lim_{n_c \to \infty}\widehat{\mathcal{D}}_{\widetilde{B}} \,=\, 
 \mathrm{E}_C\{ \mathcal{D}_B(\tau) \}
\end{equation}
where $\mathrm{E}_C\{\cdot\}$ denotes statistical expectation with respect to the distribution of $C$, and $\mathcal{D}_B(\tau)$ is given by (21). In the following, the resulting process $\widetilde{B}(t)$ will be referred to as the \emph{random-chip Bernoulli process}.

The modulation of the chip period can be viewed as a means to spread the two impulse functions, located at $t_C$ and $-t_C$, in the second derivative $R''_B(\tau)$ of the autocorrelation function $R_B(\tau)$ of the underlying synchronous Bernoulli process $B(t)$; the effects of the random modulation are illustrated schematically in Fig.\,4\,c and Fig.\,4\,d. 

As seen, the impulse functions, $\delta(\tau - t_C)$ and $\delta(\tau + t_C)$, have been transformed into respective probability density function, $p^+_C(\tau)  =  p_C(\tau-t_C)$ 
 and $p^-_C(\tau)  =  p_C(\tau+t_C)$.

\subsubsection*{Autocorrelation and Power Spectrum of $\widetilde{B}(t)$}
When $\tau\mspace{-1mu} >\mspace{-1mu} 0$, the derivative of the autocorrelation function $R_{\widetilde{B}}(\tau)$ of the random-chip Bernoulli process $\widetilde{B}(t)$ can be determined from
\begin{eqnarray}
R'_{\widetilde{B}}(\tau) &\mspace{-7mu}=\mspace{-7mu}& 
\mathrm{E}_C\{ R'_B(\tau) \}\nonumber \\
&\mspace{-7mu}=\mspace{-7mu}& 
\frac{1}{t_C} \mspace{-1mu} \int_0^\infty \mspace{-3mu}
[u(\tau-\xi) - u(\tau)] \mspace{2mu} 
p^+_C(\xi) \mspace{2mu} d\mspace{1mu}\xi \nonumber \\ 
  &\mspace{-7mu}=\mspace{-7mu}& 
  \left [F^+_C(\tau) - 1 \right ]\mspace{-1mu} /t_C, \quad \tau > 0
\end{eqnarray}
where $F^+_C(\tau)$ is the cumulative distribution function (cdf) of the rv $C$, defined by
\begin{equation}
F^+_C(\xi) \mspace{3mu}\triangleq \mspace{3mu}
\Pr \mspace{1mu}\{\mspace{1mu}C \leq \xi\mspace{1mu}\} \mspace{1mu}.
\end{equation}
Since the derivative $R'_{\widetilde{B}}(\tau)$ is an odd function of $\tau$,
\begin{equation}
R'_{\widetilde{B}}(\tau) \, = \,\mathrm {sgn}(\tau)\mspace{1mu}
\left [F^+_C(|\tau|) - 1 \right ]\mspace{-1mu} /t_C .
\end{equation}

When $\tau \mspace{-1mu} > \mspace{-1mu}0$, the autocorrelation function $R_{\widetilde{B}}(\tau)$ of the random-chip Bernoulli process $\widetilde{B}(t)$ can be expressed as
\begin{eqnarray}
R_{\widetilde{B}}(\tau) &\mspace{-11mu}=\mspace{-11mu}& 
R_{\widetilde{B}}(0) \mspace{1mu} +\mspace{1mu} 
\int_0^\tau \mspace{-8mu} R'_{\widetilde{B}}(\xi)
\mspace{2mu} d\mspace{1mu}\xi \nonumber \\
&\mspace{-11mu}=\mspace{-10mu}& 
1 \mspace{1mu} -\mspace{1mu} 
\frac{\tau}{t_C} \mspace{1mu} +\mspace{1mu} \frac{1}{t_C}\mspace{-2mu}
\int_0^\tau \mspace{-8mu}F^+_C(\xi)
\mspace{2mu} d\mspace{1mu}\xi \mspace{1mu}, \quad \tau > 0.
\end{eqnarray}
Since the autocorrelation function $R_{\widetilde{B}}(\tau)$ is an even function of $\tau$, 
\begin{equation}
R_{\widetilde{B}}(\tau) \,=\, 
1 \mspace{1mu} -\mspace{1mu} 
\frac{|\tau|}{t_C} \mspace{1mu} +\mspace{1mu} \frac{1}{t_C}\mspace{-2mu}
\int_0^{|\tau|} \mspace{-8mu}F^+_C(\xi)
\mspace{2mu} d\mspace{1mu}\xi \mspace{1mu}.
\end{equation}

From the properties of the cumulative distribution function $F^+_C(\tau)$, it follows that the autocorrelation function $R_{\widetilde{B}}(\tau)$ of a random-chip Bernoulli process $\widetilde{B}(t)$: is non-negative; remains above the tangent lines $(1\mspace{-1mu}-\mspace{-1mu}|\tau|/t_C)$; is decreasing monotonically from its maximal value, $R_{\widetilde{B}}(0)=1$, to zero, reached at $|\tau| = (1\mspace{-1mu}+\mspace{-1mu}\alpha)\mspace{1mu}t_C$. 

The power spectrum density $S_{\widetilde{B}}(\omega)$ of a random-chip Bernoulli process $\widetilde{B}(t)$ can be determined from
\begin{equation}
S_{{\widetilde{B}}}(\omega) \,=\, \mathcal {F} \{R_{\widetilde{B}}(\tau)\}
\,=\, -\frac {1} {\omega^2}\mspace{2mu} 
\mathcal {F} \{R''_{\widetilde{B}}(\tau) \}.
\end{equation}
Since
\begin{equation}
R''_{\widetilde{B}}(\tau) \,=\, [-2\mspace{1mu} \delta(0) + p^-_C(\tau) +
p^+_C(\tau)]/t_C
\end{equation}
the power spectrum density $S_{\widetilde{B}}(\omega)$ of a process $\widetilde{B}(t)$ can be expressed as
\begin{equation}
S_{{\widetilde{B}}}(\omega) 
\,=\, \frac {2} {\omega^2\mspace{1mu} t_C}\mspace{1mu} 
\left[1 - \Psi(\omega) \cos(\omega \mspace{1mu}  t_C)                                   
\right ]
\end{equation}
where $\Psi(\omega) = \mathcal {F} \{p_C(\tau)\}$.

From the properties of characteristic functions{\footnote {Because $p_C(\tau)$ is an even function of $\tau$, the Fourier transform $\Psi(\omega)$ can be replaced by the characteristic function of $p_C(\tau)$.}}, it follows that $|\Psi(\omega)|\leq 1$; consequently, the power spectrum density $S_{\widetilde{B}}(\omega)$ assumes non-negative values for the entire range of $\omega$.
 
\begin{figure}[] %[b]
	\centering
	\includegraphics[width=8.8cm]{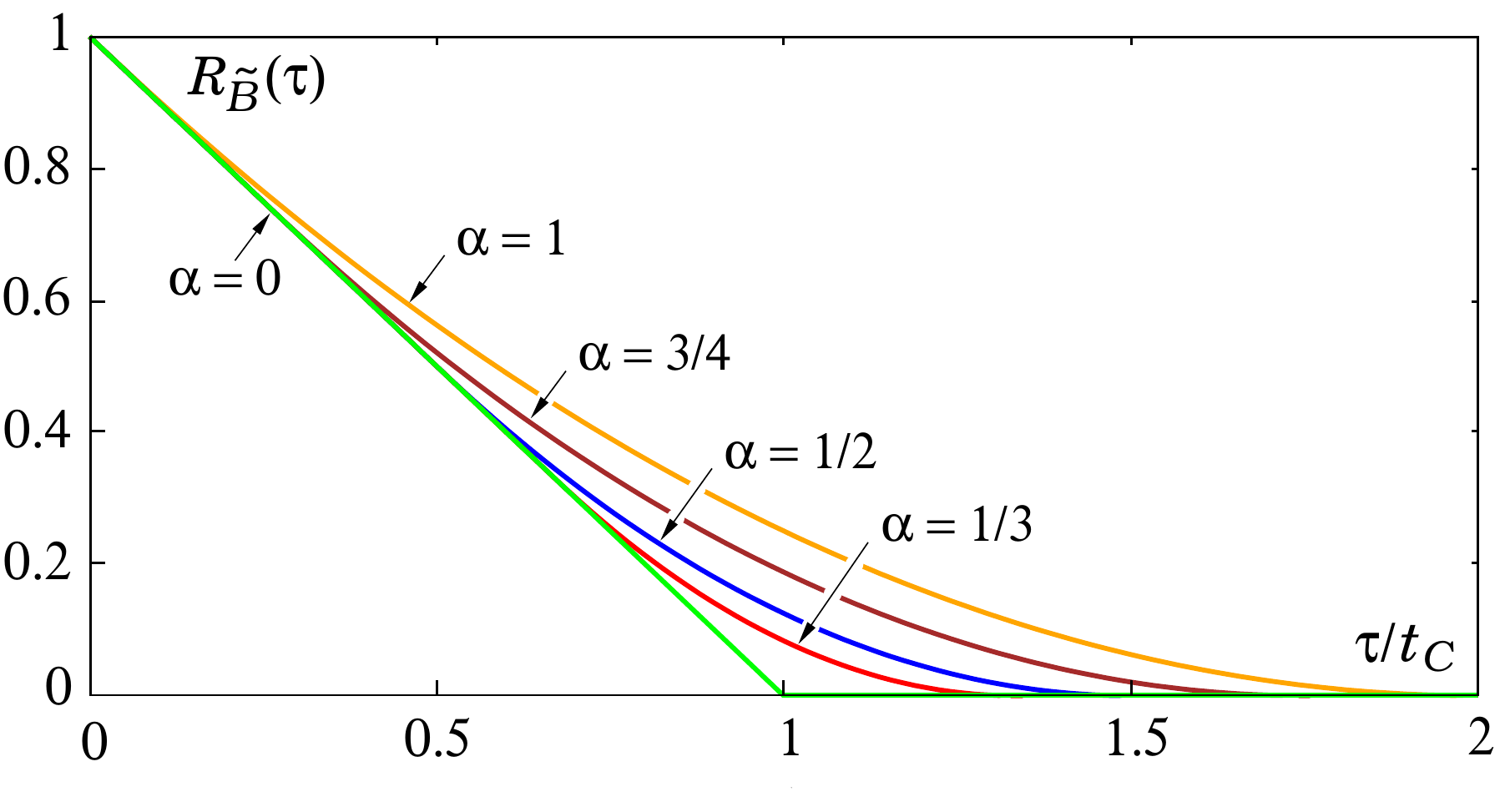}
	\vspace{-0.9cm}
	\caption{Autocorrelation function $R_{\widetilde{B}}(\tau)$ of a random-chip Bernoulli process with uniform chip modulation.}
\end{figure}

\subsection{Optimal Modulation of Chip Period}
An examination of Fig.\,3\,c and Fig.\,4\,d may lead to an intuitive conclusion that the optimal autocorrelation function could result from a uniform modulation of the chip period. To show that such a conjucture is indeed true, assume that the random chip period $C$ has a uniform distribution 
\begin{equation}
p^+_C(\tau) = 1/(2\mspace{1mu} \alpha\mspace{1mu} t_C),\quad
(1\mspace{-1mu}-\mspace{-1mu}\alpha)\mspace{1mu} t_C
< \tau <
 (1\mspace{-1mu}+\mspace{-1mu}\alpha)\mspace{1mu} t_C
\end{equation}
where $0 \leq \alpha \leq 1$.

The cumulative distribution function, $F^+_C(\tau)$, of the chip period is of the form
\begin{equation}
F^+_C(\tau)\,=\,
\left\{
\begin{array}{l@{\quad}l} \!\!\!0, & {|\tau| \le (1\mspace{-2mu}-\mspace{-2mu}\alpha)\mspace{1mu}t_C}\\
\!\!\! 1, & {|\tau| > (1\mspace{-2mu}+\mspace{-2mu}\alpha)\mspace{1mu}t_C} \\
\!\! \! {[\tau\mspace{-2mu} +\mspace{-2mu} (\alpha\mspace{-2mu}-\mspace{-2mu}1)t_C]}
\big /
{2\mspace{1mu} \alpha\mspace{1mu} t_C}, & \mbox{otherwise}\, .
\end{array} \right . 
\end{equation}
From (29) and (34) it follows that when $t_C\mspace{1mu} =\mspace{1mu} 1$,
\begin{equation}
R_{\widetilde{B}}(\tau) \,=\, \widetilde{\Lambda}(\tau;\alpha) .
\end{equation}

For illustration purposes, Fig.\,5 shows plots of the autocorrelation function $R_{\widetilde{B}}(\tau)$ of a Bernoulli process with uniform chip modulation for selected values of the modulation parameter $\alpha$. If there is no modulation ($\alpha =0$), then $R_{\widetilde{B}}(\tau)$ assumes the form of 
$R_{\widetilde{B}}(\tau) = {\Lambda}\mspace{1mu} (\tau; t_C)$.

\begin{figure}[] 
	\centering
	\includegraphics[width=8.95cm]{sza6.pdf}
	\vspace{-0.9cm}
	\caption{Power spectrum density $S_{{\widetilde{B}}}(\omega)$ of a random-chip Bernoulli process with uniform chip modulation.}
\end{figure}

In this case,
$$
\Psi(\omega) = \mathcal {F} \{p_C(\tau)\} = {\sin(\omega  \alpha\mspace{1mu} t_C) }
\big / \mspace{1mu}
   {(\mspace{1mu}\omega  \alpha\mspace{1mu} t_C)}
$$
and the power spectrum density assumes the form
\begin{equation}
S_{{\widetilde{B}}}(\omega) \,=\,
\frac {\mspace{-3mu}2}{\omega^2\mspace{1mu} t_C} \mspace{-3mu}
\left[1 \mspace{-1mu}-\mspace{-1mu}
\frac {\sin(\omega  \alpha\mspace{1mu} t_C)\cos(\omega t_C) }{\omega  \alpha\mspace{1mu} t_C} 
\right] \mspace{-4mu},\quad |\omega| > 0
\end{equation}
and when $\omega \to 0$, $S_{{\widetilde{B}}}(\omega) \to 1\mspace{-1mu}+\mspace{-1mu} {\alpha^2\mspace{-1mu}}/{3}$.

If there is no modulation ($\alpha =0$), then $S_{{\widetilde{B}}}(\omega)$ converges to the psd (15) of a constant-chip Bernoulli process $B(t)$.

Fig.\,6 shows plots of the power spectrum density $S_{{\widetilde{B}}}(\omega)$ of a Bernoulli process with uniform chip modulation for selected values of the modulation parameter $\alpha$.

 \subsubsection*{Other Modulation Distributions}
As shown above, the use of uniform distribution for chip modulation has resulted in the postulated optimum shape (20) of the autocorrelation function $R_{{\widetilde{B}}}(\tau)$, However, for such purpose, it is also possible to exploit other distributions that are unimodal and even functions of $\tau$  and have a finite support.

As an instructive example, consider a raised cosine distribution,
\begin{eqnarray}
p_C(\tau) =
\frac {1}{2\mspace{1mu} \alpha\mspace{1mu} t_C} \mspace{-3mu} \left [
1 \mspace{-1mu}+ \mspace{-1mu} \cos \mspace{-2mu}
\left( \frac {\pi\mspace{1mu} \tau}{\alpha \mspace{1mu}t_C} \right )
\mspace{-1mu} \right]  =  
\frac {1}{\alpha\mspace{1mu} t_C} \mspace{-1mu}
\cos^2\mspace{-3mu} \left(\mspace{-1mu} \frac {\pi\mspace{1mu} \tau}
{2\mspace{1mu} \mspace{1mu} \alpha\mspace{1mu} t_C} \mspace{-1mu}\right )
\\
|\tau| \leq \alpha\mspace{1mu} t_C,  \quad
0 < \alpha \leq 1  \mspace{1mu}. \nonumber
\end{eqnarray}
When $\alpha\mspace{-1mu} \to\mspace{-1mu} 0$, the resulting autocorrelation function $R_{\widetilde{B}}(\tau)$ tends to the triangular function (13), whereas, for $\alpha=1$, the autocorrelation function $R_{\widetilde{B}}(\tau)$ assumes the form
\begin{equation}
R_{{\widetilde{B}}}(\tau) \,=\,
1 - \frac{|\tau|} {t_C} + \frac{\tau^2} {4\mspace{1mu} t^2_C} 
- \frac{1}{\pi^2}\mspace{-1mu} \sin^2\mspace{-3mu} \left (\mspace{-1mu}
\frac {\pi\mspace{1mu} \tau}{2\mspace{1mu} t_C}\mspace{-1mu}
\right ) \mspace{-2mu} .
\end{equation}

Fig.\,7 shows a plot of the function (38) along with the plots of two autocorrelation functions resulting from uniform chip modulation with respective parameter values: $\alpha\!=\!0$ (no modulation) and $\alpha\!=\!1$ (maximal spread).
 
The uniform chip modulation establishes both a lower bound ($\alpha\!=\!0$) and an upper bound ($\alpha\!=\!0$) on the autocorrelation function $R_{{\widetilde{B}}}(\tau)$ of a random-chip Bernoulli process, when the  modulation distribution is a unimodal and even function of $\tau$ with a finite support. From this viewpoint, the uniform distribution may be regarded as an optimal choice.
 
However, the condition of unimodality of the modulation distribution has been imposed for mathematical convenience. An even, but not necessarily unimodal, distribution with finite support 
can also be employed for random chip modulation.

\subsection{A Product of Two Random Binary Processes}
 A random binary process can also be obtained from a product of two, or more, random binary processes [16]. If at least one of the processes is a random-chip Bernoulli process $R_{{\widetilde{B}}}(\tau)$ discussed above, then the resulting random binary process will have an impulse-like autocorrelation function.

Let $V_1(t)$ and $V_2(t)$ be two uncorrelated random binary processes with respective autocorrelation functions, $R_{V1}(\tau)$ and $R_{V2}(\tau)$, and zero-crossing rates, $n_1$ and $n_2$. 
Consider the product, 
\begin{equation}
Z(t)=V_1(t)\mspace{1mu} V_2(t)
\end{equation}
of the two processes. By construction, $Z(t)$ is also a random binary process, and its autocorrelation function is given by 
\begin{equation}
R_Z(\tau) = R_{V1}(\tau)\mspace{1mu} R_{V2}(\tau).
\end{equation}
The zero-crossing rate $n_Z$ of the random process $Z(t)$ can be determined from
\begin{equation}
n_Z = 
 -\mspace{-2mu}\lim_{\tau \to 0^+} \frac {\partial}{2\mspace{1mu} \partial \mspace{1mu} \tau} 
 \left[R_{V1}(\tau)\mspace{1mu} R_{V2}(\tau) \right] = n_1 + n_2.
\end{equation}

Therefore, a random binary process with optimal correlation properties can be obtained by multiplying a random-chip Bernoulli process ${\widetilde{B}}(t)$ by another binary process $V(t)$. Since the autocorrelation function $R_{\widetilde{B}}(\tau)$ vanishes for $|\tau| >$ $(1+ \alpha \mspace{1mu} t_C)$, the autocorrelation function $R_Z(\tau)$ of the resulting product will always have a finite support. The process $V(t)$ may be of (pseudo)random, chaotic or even deterministic nature, and the resulting power spectral density $S_Z(\omega)$ will be the convolution of the two respective densities, $S_{\widetilde{B}}(\omega)$ and 
$S_V(\omega)$.

It is important to note that either of the two underlying random processes, $V_1(t)$ and $V_2(t)$, can be recovered from the product process $Z(t)$ by making use of the 'demodulation' property
\begin{eqnarray}
Z(t)\mspace{1mu} V_2(t) \,=\, V_1(t)\mspace{1mu} [V_2(t)]^2 \,=\, V_1(t)\mspace{6mu}&& \nonumber \\
Z(t)\mspace{1mu} V_1(t) \,=\, V_2(t)\mspace{1mu} [V_1(t)]^2 \,=\, V_2(t)\mspace{1mu}.&&
\end{eqnarray}
While the above operation has been exploited in conventional (discrete-time and synchronous) spread-spectrum systems, it may also find a much wider application in developing a broad class of continuous-time techniques for reliable/covert transmission of binary data.

\begin{figure}[] %[b]
	\centering
	\includegraphics[width=8.cm]{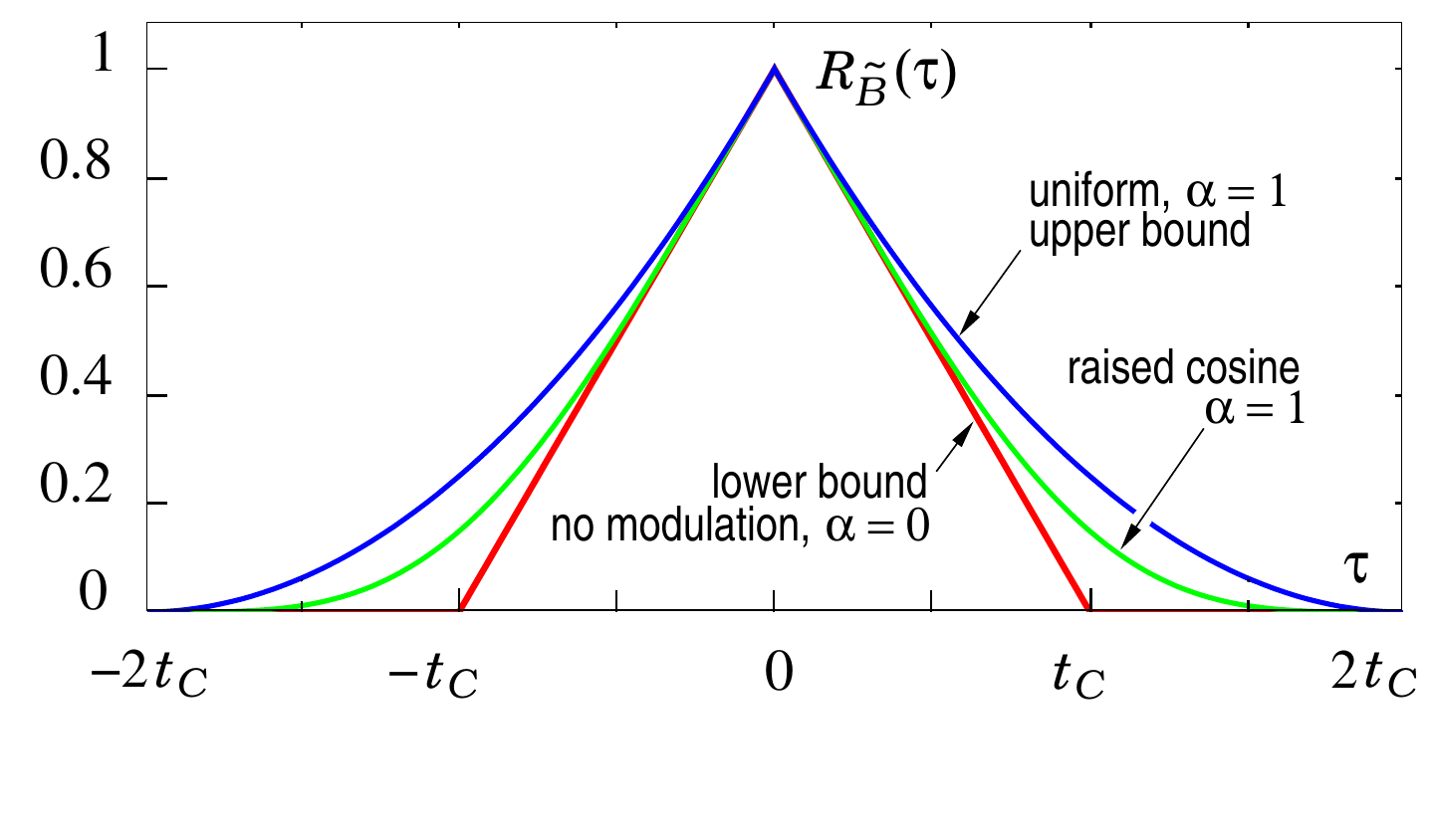}
	\vspace{-1.0cm}
	\caption{Bounds on the autocorrelation function $R_{{\widetilde{B}}}(\tau)$ of a random-chip Bernoulli process.}
\end{figure} 

\subsection{Basic Implementation Techniques}
The generation of a random-chip Bernoulli process involves two random mechanisms: one, to generate a constant-chip Bernoulli process and another one, to randomly modulate the chip duration. In practice, either of the two (or both) random mechanisms can be replaced by a pseudorandom one.
\begin{figure}%[b]
	\begin{center}
		\includegraphics[width=6.9cm]{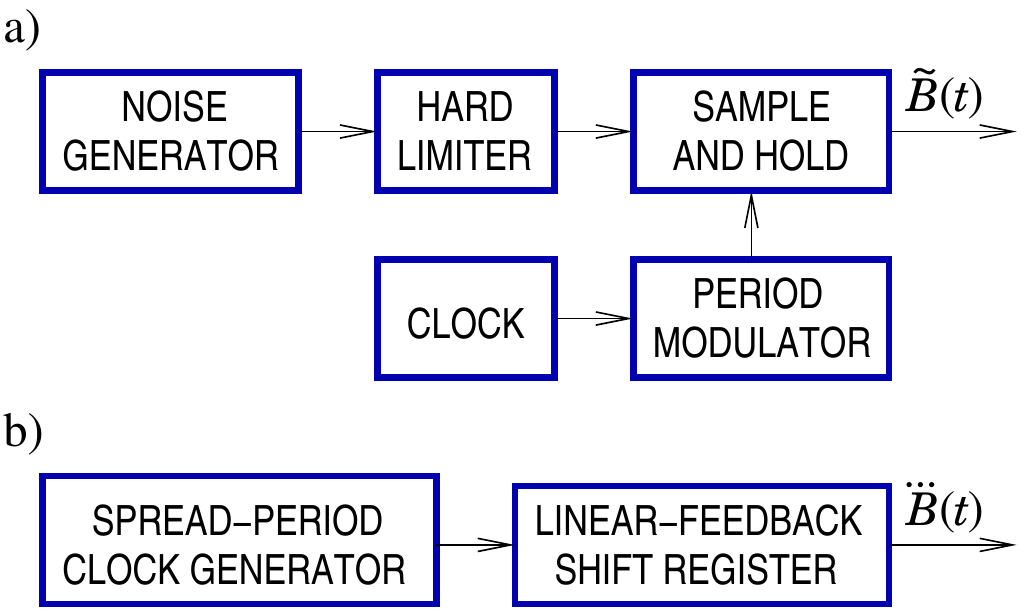}
	\end{center}
	\vspace{-0.3cm}
	\caption{Generation of$\mspace{1mu}$: (a) a random Bernoulli process $\widetilde{B}(t)$ with random chip modulation; (b) a pseudorandom Bernoulli process $\dddot{B}(t)$ with pseudorandom chip modulation.}
\end{figure}

Fig.\,8\,a is a block diagram of a system that exploits two random mechanisms. 

The {\sf {\scriptsize {NOISE GENERATOR}}} supplies a random signal with suitable characteristics. When there is no random period modulation, the output of the {\sf {\scriptsize {HARD LIMITER}}} is sampled at constant intervals, determined by the {\sf {\scriptsize {CLOCK}}} period, and the {\sf {\scriptsize {SAMPLE-AND-HOLD}}} produces a constant-chip Bernoulli process ${B}(t)$. 

However, when the {\sf {\scriptsize {PERIOD MODULATOR}}} randomly perturbs the constant period duration, sampling of the 
output of the {\sf {\scriptsize {HARD LIMITER}}} becomes randomly irregular, and the {\sf {\scriptsize {SAMPLE-AND-HOLD}}} produces a random-chip Bernoulli process  $\widetilde{B}(t)$. 

Fig.\,8\,b is a simplified block diagram of a system that exploits two pseudorandom mechanisms. 

The {\sf {\scriptsize {SPREAD-PERIOD CLOCK GENERATOR}}} supplies a sequence of a predetermined number of pulses, and the interpulse intervals are selected in a pseudorandom fashion from a prescribed set of deterministic values. Such formed pulse sequence may then be repeated cyclically, or each new cycle may use a different permutation of the interpulse intervals [17]. The resulting infinite pulse sequence is driving the {\sf {\scriptsize {LINEAR-FEEDBACK SHIFT REGISTER}}} operating in a standard configuration [18]. 

When a constant-frequency clock is used, such a configuration is known to generate a pseudorandom binary waveform that may be regarded as a realization of a constant-chip Bernoulli process. However, in the arrangement shown in Fig.\,8\,b, the {\sf {\scriptsize {SPREAD-PERIOD CLOCK GENERATOR}}} is employed, and the {\sf {\scriptsize {LINEAR-FEEDBACK SHIFT REGISTER}}} produces a representation of a pseudorandom Bernoulli process with pseudorandom chip modulation $\dddot{B}(t)$ [19].

In practical hardware implementations, it may be more convenient to employ a voltage-controlled oscillator (VCO) to modulate the clock period. In such a case, to find a required voltage distribution, a specified voltage-frequency characteristic will have to be converted into a voltage-period relationship{\footnote {Linear period modulation is equivalent to hyperbolic frequency modulation or logarithmic phase modulation.}} [20].

\section{Conclusions}
It has been shown that a random binary process with impulse-like autocorrelation  can be generated by random or pseudorandom sampling of the output of a source that supplies a symmetric random binary waveform. Therefore, changing the sampling pattern, yet preserving its statistical characteristics, will result in an uncorrelated version of the underlying binary waveform being sampled.

The techniques presented in this paper are particularly well suited to hardware implementation utilizing standard analog/digital building blocks. This aspect is very important when developing a multi-user system to be employed in mass-produced units, such as automotive radar or autonomous radio-frequency sensors.

\section*{Acknowledgment}
The author gratefully acknowledges the many stimulating discussions with Prof. Miroslaw Bober on statistical signal and image processing.

\end{document}